\documentclass{article}

\usepackage{epsfig}
\usepackage{graphics}
\usepackage{graphicx}

\usepackage[]{algorithm2e}
\usepackage{longtable}
\usepackage{multirow}
\usepackage{url}
\usepackage{subfigure}
\usepackage{hyperref}

\begin{document}

\title{Universal One-Dimensional Cellular Automata Derived for Turing Machines and its Dynamical Behaviour\footnote{{\em International Journal of Unconventional Computing}. \url{https://www.oldcitypublishing.com/journals/ijuc-home/ijuc-issue-contents/ijuc-volume-14-number-2-2019/ijuc-14-2-p-121-138/}}}

\author{Sergio J. Mart{\'i}nez$^{1}$, Iv\'an M. Mendoza$^{1}$ \\ Genaro J. Mart{\'i}nez$^{1,2}$, Shigeru Ninagawa$^{3}$}


\maketitle

\begin{centering}
$^1$ Laboratorio de Ciencias de la Computaci\'on, Escuela Superior de C\'omputo, Instituto Polit\'ecnico Nacional, M\'exico \\
\url{serj1495@gmail.com}, \url{manzanoivan95@gmail.com} \\
$^2$ Unconventional Computing Lab, University of the West of England, Bristol, United Kingdom. \\
\url{genaro.martinez@uwe.ac.uk} \\
$^3$ Kanazawa Institute of Technology, Hakusan, Japan \\ 
\url{ninagawa@infor.kanazawa-it.ac.jp} \\
\end{centering}

\begin{abstract}
Universality in cellular automata theory is a central problem studied and developed from their origins by John von Neumann. In this paper, we present an algorithm where any Turing machine can be converted to one-dimensional cellular automaton with a 2-linear time and display its spatial dynamics. Three particular Turing machines are converted in three universal one-dimensional cellular automata, they are: binary sum, rule 110 and a universal reversible Turing machine.
\end{abstract}


\section{Introduction}

In this paper, we research the spacial dynamics of Turing machines as discrete dynamical systems and its conversion as cellular automata. Cellular automata are dynamical systems that evolve in states and time discrete. Historically cellular automata are used recurrently to explore novel unconventional domains of computability across of signals, gliders, particles, waves, and mobile self-localization interactions \cite{kn:Hey98, kn:MTV86, kn:Wolf88, kn:Ada02, kn:MAS11, kn:MAM18, kn:MM18, kn:Zen12}.

Turing machines are computable models invented by Alan Turing in 1936 to proof if a problem has a solution or not \cite{kn:Tur36, kn:Sha56, kn:Dav65}. The behaviour of such machine is displayed as a sequence of instantaneous descriptions, which describes a string on an infinite tape where every instantaneous description is determined by the sequential selection of a rule from a finite set of rules. If such sequence of instantaneous descriptions reaches a final state or cannot do more descriptions hence the computation is done.

Frequently the construction of computers in cellular automata are handled just with primitive signals (classic examples include von Neumann rule \cite{kn:von66}, Codd \cite{kn:Codd68}, Banks \cite{kn:Ban71}) or with non-trivial patterns known as gliders, particles, waves or self-localizations (classic examples include the Game of Life \cite{kn:Gard70}, Rule 110 \cite{kn:Cook04, kn:Wolf02}), and recently by packages of signals (\cite{kn:MMA10, kn:MAM10}). Actually, we have some Turing machines implemented in cellular automata, one of them in the Game of Life by Paul Rendell \cite{kn:Ren16} and in one dimension by Smith \cite{kn:Smi71} and other by Lindgren and Nordahl \cite{kn:LN90}.

The paper has the next structure. Initially, we introduce basic concepts to Turing machines and cellular automata. Later, we show an algorithm able to convert a Turing machine to its equivalent one-dimensional cellular automaton. Finally, we show conversions of three particular Turing machines to three one-dimensional cellular automata.

\section{Basic concepts}

\subsection{Turing machines}
A Turing machine is defined as a 7-tuple $M = (Q,\Sigma,\Gamma,\delta,$ $q_0,B,F)$, where $Q$ is the finite set of states, $\Sigma$ the alphabet of input symbols, $\Gamma$ is the complete set of symbols in the tape, $\delta$ the transition function, $q_0$ the start state, $B$ the blank symbol and $F$ the set of final or accepting states. The transition function determines relations between a state with a symbol to other state, as follows: $\delta(q,X) \rightarrow (p,Y,D)$ $\forall$ $q,p \in Q$ and $X,Y \in \Gamma$, and a direction $D$ (left or right). This way, the language that a Turing machine can recognize is defined as $L(M) = \{w \,\,|\,\,q_0w\vdash^*I \,\,\forall$ $I \in F\}$, where $w$ is a string of $L$ and $\vdash$ denotes a move of a Turing machine \cite{kn:HU79, kn:Mins67}.

\subsection{Cellular automata}
One-dimensional cellular automata is represented by an array of {\it cells} $x_i$ where $i \in Z$ and each $x$ takes a value from a finite alphabet $\Sigma$. Thus, a sequence of cells \{$x_i$\} of finite length $n$ describes a string or {\it global configuration} $c$ on $\Sigma$. The set of finite configurations will be expressed as $\Sigma^n$. An evolution is comprised by a sequence of configurations $\{c_i\}$ produced by the mapping $\Phi:\Sigma^n \rightarrow \Sigma^n$; thus the global relation is symbolized as: $\Phi(c^t) \rightarrow c^{t+1}$, where $t$ represents time and every global state of $c$ is defined by a sequence of cell states. The global relation is determined over the cell states in configuration $c^t$ updated at the next configuration $c^{t+1}$ simultaneously by a local function $\varphi$ as follows: $\varphi(x_{i-r}^t, \ldots, x_{i}^t, \ldots, x_{i+r}^t) \rightarrow x_i^{t+1}$. Wolfram represents one-dimensional cellular automata with two parameters $(k,r)$, where $k = |\Sigma|$ is the number of states, and $r$ is the neighbourhood radius. In this domain, we have that $\Sigma^n$ determines the number of neighbourhoods of size $n=2r+1$, and $k^{k^n}$ represents the number of distinct evolution rules \cite{kn:Wolf84, kn:Mc09}.

\section{Converting a Turing machine to a one-dimensional cellular automaton}
In 1971 \cite{kn:Smi71}, Alvy Smith III proved that a Turing machine of $m$ internal states and $n$ tape symbols in 2-linear time can be simulated by a one-dimensional cellular automaton with a neighbourhood of radius $r = 1$ and $m+2n$ colours (cellular automata states). Later in 1990 \cite{kn:LN90}, Lindgren and Nordahl pointed that not all $m+2n$ colours is necessary to accomplish this and asserted that $m + |\{(state, direction) \, \mid \, \exists \, \delta(p,X) = (state, X', direction)\}|$ colours are necessary. It is $m$ plus the cardinality of the set of pairs $(state, direction)$ such that it is possible for the machine to reach such state by moving in such direction. Nevertheless, this number of colours is not enough to do the simulation according to the idea shown by Smith, since there are needed at least 2 colours to represent the machine's head when it moves to the left. This way, the smaller number of colours required to accomplish this particular simulation is $m + n + |\{ state \, \mid \, \exists \, \delta(p,X) = (state, X', left)\}|$ ($m$ plus $n$ plus the cardinality of the set of states that can be reached by moving to the left).

\subsection{Algorithm construction}
To an arbitrary Turing machine $M = (Q, \Sigma_M, \Gamma, \delta, q_0, B, F)$ there exists a one-dimensional equivalent cellular automaton $C$ with an alphabet $\Sigma_C$ and radius $r = 1$. So, it is constructed from \cite{kn:LN90}, as following.

$\Sigma_C = \Sigma_M \cup Q \cup Q'$, where $Q'$ is a set of auxiliary symbols that will help to simulate the movements of the head to the left. So, the number of auxiliary symbols for the cellular automaton will be $|Q'| \leq |Q|$. Note that $\Sigma_C$ is formed by the union of a set of tape symbols $\Sigma_M$ and a set of states $Q$. This is possible because they are both treaties just as sets of symbols and not as symbols and states separately.

The movements to the right $\delta(p, X) = (q, Y, right)$ for a cellular automaton is defined in the next way:

$\varphi(Z_1,   p,  X ) \rightarrow Y$.

$\varphi(p,   X, Z_2 ) \rightarrow q$.

\noindent where $n$ is a natural number on $Z_n$ and indicates every symbol on $\Sigma_C$. That means that for every rule shown, there are defined $|\Sigma_C|$ rules on $\varphi$. By applying this rules, we get a behaviour like this,

\begin{table}[th]
\centering
\caption{Simulation of the head movements to the right}
\begin{tabular}{|c|c|c|c|}
\hline
$\sigma_1$ & $p$ & $X$ & $\sigma_2$ \\ \hline
$\sigma_1$ & $Y$ & $q$ & $\sigma_2$ \\ \hline
\end{tabular}
\end{table}

\noindent where $\sigma_1, \sigma_2 \in \Sigma_M$. Finally, for the Turing machine's transitions to the left, $\delta(p, X) = (q, Y, left)$, we define the following rules on $\varphi$.

$\varphi(Z_1,   p,  X ) \rightarrow q'$.

$\varphi(p,   X, Z_2 ) \rightarrow Y$.

$q' \in Q'$ is a new symbol and it is an auxiliary symbol which indicates that in the next time step the head of the machine will move to the left. To simulate this one, we need add the next rule.

$\varphi(Z_1,  Z_2, q' ) \rightarrow q$.

$\varphi(Z_1,  q', Z_2 ) \rightarrow Z_1$.

By applying these rules, we get a behaviour like this ($\sigma_1, \sigma_2 \in \Sigma_M$):

\begin{table}[th]
\centering
\caption{Simulation of the head movements to the left}
\begin{tabular}{|c|c|c|c|}
\hline
$\sigma_1$ & $p$ & $X$ & $\sigma_2$ \\ \hline
$\sigma_1$ & $q'$ & $Y$ & $\sigma_2$ \\ \hline
$q$ & $\sigma_1$ & $Y$ & $\sigma_2$ \\ \hline
\end{tabular}
\end{table}

Note that the simulation of a Turing machine by using these rules will take at most twice the time used by the Turing machine itself.

\subsection{Algorithm description}

By generalizing the sets of rules shown, we can build an algorithm such that, given a Turing machine $M$ as input produces an equivalent cellular automaton $C$. 

    \begin{algorithm}
        \KwData{Turing machine, $M = (Q, \Sigma_M, \Gamma, \delta, q_0, B, F)$}
        \KwResult{One-dimensional cellular automaton, $C$}
        $\Sigma_C = \Sigma_M \cup Q$\;
        $\varphi = \emptyset$\;
        \For{every $p \in Q, X \in \Sigma_M$ such that $\delta(p, X) = (q, Y, D)$}{
            \If{D = Right}{
               Add $Z_1pX \rightarrow Y$ to $\varphi$\;
               Add $pXZ_1 \rightarrow q$ to $\varphi$\;
            }
            \If{D = Left}{
                \If{$q' \notin \Sigma_C$}{
                    Add $q'$ to $\Sigma_C$\;
                    Add $Z_1Z_2q' \rightarrow q$ to $\varphi$\;
                    Add $Z_1q'Z_2 \rightarrow Z_1$ to $\varphi$\;
                }
                Add $Z_1pX \rightarrow q'$ to $\varphi$\;
                Add $pXZ_1 \rightarrow Y$ to $\varphi$\;
            }
        }
       \For{every undefined $\varphi(W_1, W_2, W_3)$, $W_1, W_2, W_3 \in \Sigma_C$}{
            Add $W_1W_2W_3 \rightarrow W_2$ to $\varphi$\;
       }
    \end{algorithm}

Note that the cycle at the end of the algorithm that takes $O(n^3)$ time, where $n = |\Sigma_C|$. It is only used to indicate that every undefined combination of symbols will keep constant. By taking this for granted, the algorithm takes just $O(m*o)$ time where $m = |Q|$ and $o = |\Sigma_M|$.

\begin{figure}
\centering
\includegraphics[width=0.68\textwidth]{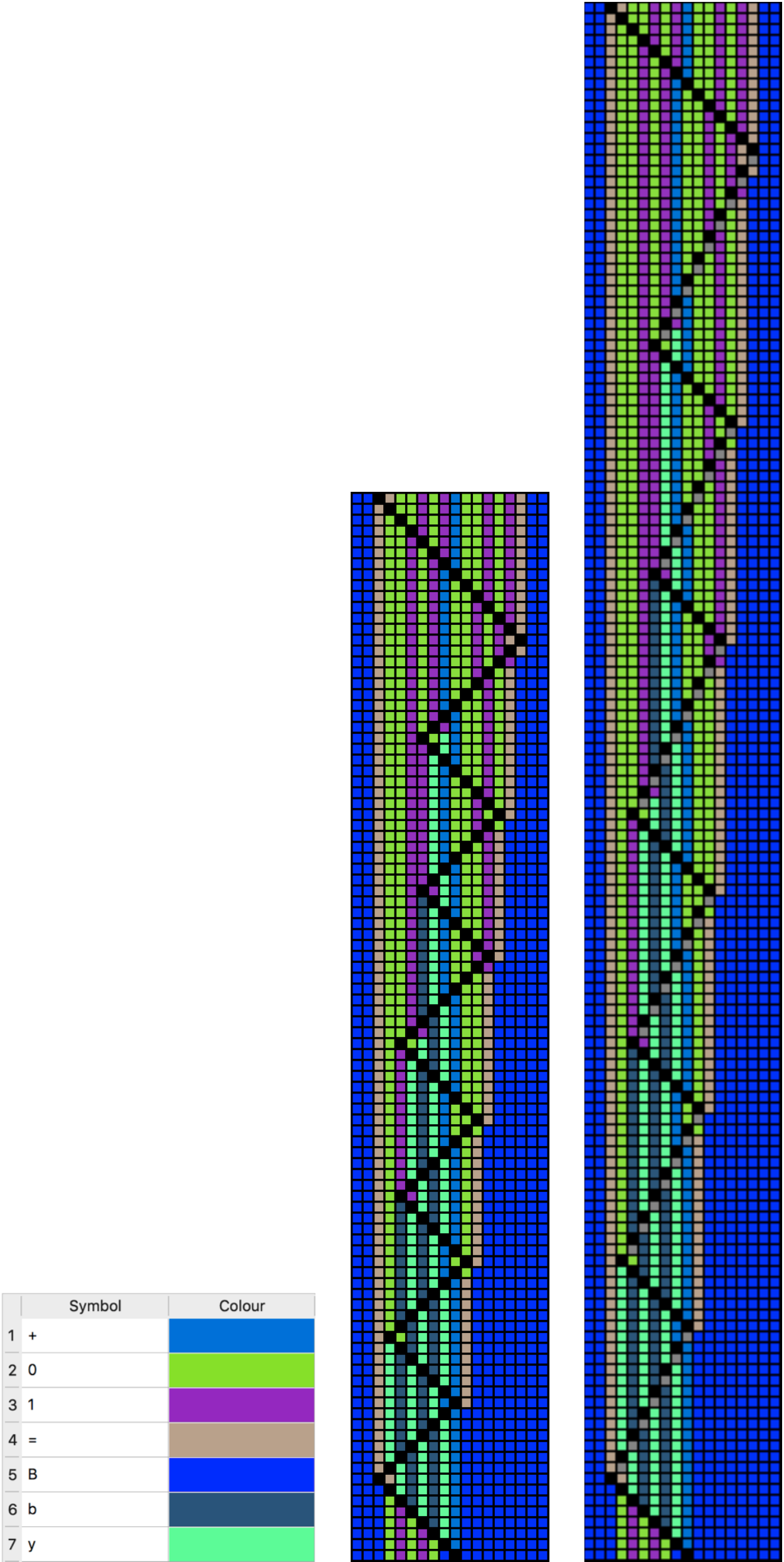}
\caption{Behaviour of the Turing machine $M$ and its equivalent cellular automaton $C_{BS}$ for the input string $=00101+00101=$. (left) Representation of the symbols. (center) Turing machine dynamics. (right) Cellular automaton dynamics.}
\label{fig:exec04}
\end{figure}

\subsection{Simulation ending}

The halting condition is well specified for Turing machines but not for cellular automata. For a cellular automaton $C$ the simulation of a Turing machine ends when the global configuration $w$ remains constant. This way, the automaton has finished the analysis of the string $\omega_0$ at a time step $t$ when $\omega_t = \omega_{t+1}$.
    
Similarly, we say the string $\omega_0$ has been accepted by $C$ if $\omega_t = \alpha f\beta$ where $f \in F$, once $C$ has ended the simulation at a time step $t$.

\section{Universal cellular automata from Turing machines}

\subsection{Binary sum}

A classic example in computer science is the design of computable functions performing arithmetical operations. In this way, a Turing machine performing sum is our first case of study.

Let $M = (Q_M, \Sigma_M, \Gamma, \delta, q_0, B, F)$ be a Turing machine capable of recognizing strings of the form $=a+b=$, where $a,b$ are binary numbers and returning $a+b$ (this machine was presented in \cite{kn:Mins67, kn:MC90}). $Q_M = \{a, Q0, dig, zle,$ $ole, zad, oad, car, new, rig, lef, fin\}$, $\Sigma_M = \{=,1,0,+\}$, $\Gamma = \{=,1,0,+,$ $y,b,B\}$ and $\delta$ is the transition function shown in the Table~\ref{TMBS}. $q_0 = a$ is the initial state of the machine and $B$ is the blank symbol.

\begin{table*}[th]
\centering
\caption{Transition table for the binary sum Turing machine $(13,7)$.}
\label{TMBS}
\tiny
\begin{tabular}{c|c|c|c|c|c|c|c}
        State & $=$ & $1$ & $0$ & $+$ & $B$ & $y$ & $b$ \\
        \hline
        $a$ & $a,=,R$ & $a,1,R$ & $a,0,R$ & $a,+,R$ & $Q0,B,L$ & \\ 
        $Q0$ & $dig,B,L$ &  &  &  &  & \\
        $dig$ &  & $ole,=,L$ & $zle,=,L$ & $lef,+,L$ &  & \\
        $zle$ &  & $zle,1,L$ & $zle,0,L$ & $zad,+,L$ &  &  & \\
        $ole$ &  & $ole,1,L$ & $ole,0,L$ & $oad,+,L$ &  &  & \\
        $zad$ & $new,y,L$& $rig,b,R$ & $rig,y,R$ &  &  & $zad,y,L$ & $zad,b,L$ \\
        $oad$ & $new,b,L$ & $car,y,L$ & $rig,B,R$ &  &  & $oad,y,L$ & $oad,b,L$ \\
        $car$ & $new,1,L$ & $car,0,L$ & $rig,1,R$ &  &  &  & \\
        $new$ &  &  &  &  & $rig,=,R$ &  & \\
        $rig$ & $dig,B,L$ & $rig,1,R$ & $rig,0,R$ & $rig,+,R$ &  & $rig,y,R$ & $rig,b,R$ \\
        $lef$ & $fin,B,R$ & $lef,1,L$ & $lef,0,L$ &  &  & $lef,y,L$ & $lef,b,L$ \\
        $fin$ &  & $fin,1,R$ & $fin,0,R$ & $halt,B,R$ &  & $fin,0,R$ & $fin,1,R$ \\
        $halt$ &  &  &  &  &  &  & \\
\end{tabular}
\end{table*}

The behaviour of the Turing machine $M$ for the input string $=00101+00101=$, is shown in Fig.~\ref{TMBS}(center). By using $M$ as input for the algorithm described, we get a cellular automaton $C_{BS}$, thich behaves almost as $M$. This is shown in Fig.~\ref{TMBS}(right). As we can see, the inputs and outputs of both $M$ and $C_{BS}$ are the same, that means the simulation was done properly. However, the behaviour of both models is not exactly alike; $C_{BS}$ takes twice the time of $M$ to move its head to the left. The set of rules defined for $C_{BS}$ is shown in Table~\ref{cuad:specautoSuma}.

\begin{table}
\centering
\caption{Set of rules defined for $C_{BS}$ to simulate $M$.}
\label{cuad:specautoSuma}
\scriptsize
\begin{tabular}{|c|c|c|c|c|}
\hline
\multicolumn{3}{|c|}{$u(x)$} & $f(u(x))$ & equivalent $M$ transitions \\ \hline \hline
$Z_1$ & $Z_2$ & $Q0'$ & $Q0$ & - \\ \hline
$Z_1$ & $Z_2$ & $dig'$ & $dig$ & - \\ \hline
$Z_1$ & $Z_2$ & $zle'$ & $zle$ & - \\ \hline
$Z_1$ & $Z_2$ & $ole'$ & $ole$ & - \\ \hline
$Z_1$ & $Z_2$ & $zad'$ & $zad$ & - \\ \hline
$Z_1$ & $Z_2$ & $oad'$ & $oad$ & - \\ \hline
$Z_1$ & $Z_2$ & $car'$ & $car$ & - \\ \hline
$Z_1$ & $Z_2$ & $new'$ & $new$ & - \\ \hline
$Z_1$ & $Z_2$ & $lef'$ & $lef$ & - \\ \hline
$Z_1$ & $a$ & $=$ & $=$ & \multirow{2}{*}{$\delta(a,=) = (a,=,right)$} \\ \cline{1-4} $a$ & $=$ & $Z_1$ & $a$ &  \\ \hline
$Z_1$ & $a$ & 1 & 1 & \multirow{2}{*}{$\delta(a,1) = (a,1,right)$} \\ \cline{1-4} $a$ & 1 & $Z_1$ & $a$ &  \\ \hline
$Z_1$ & $a$ & 0 & 0 & \multirow{2}{*}{$\delta(a,0) = (a,0,right)$} \\ \cline{1-4} $a$ & 0 & $Z_1$ & $a$ &  \\ \hline
$Z_1$ & $a$ & $+$ & $+$ & \multirow{2}{*}{$\delta(a,+) = (a,+,right)$} \\ \cline{1-4} $a$ & $+$ & $Z_1$ & $a$ &  \\ \hline 
$Z_1$ & $a$ & $B$ & $Q0'$ & \multirow{2}{*}{$\delta(a,B) = (Q0,B,left)$} \\ \cline{1-4} $a$ & $B$ & $Z_1$ & $B$ &  \\ \hline 
$Z_1$ & $Q0$ & $=$ & $dig'$ & \multirow{2}{*}{$\delta(Q0,=) = (dig,B,left)$} \\ \cline{1-4} $Q0$ & $=$ & $Z_1$ & $B$ &  \\ \hline
$Z_1$ & $dig$ & 0 & $zle'$ & \multirow{2}{*}{$\delta(dig,0) = (zle,=,left)$} \\ \cline{1-4} $dig$ & 0 & $Z_1$ & = &  \\ \hline
$Z_1$ & $dig$ & 1 & $ole'$ & \multirow{2}{*}{$\delta(dig,1) = (ole,=,left)$} \\ \cline{1-4} $dig$ & 1 & $Z_1$ & = &  \\ \hline
$Z_1$ & $dig$ & $+$ & $lef'$ & \multirow{2}{*}{$\delta(dig,+) = (lef,+,left)$} \\ \cline{1-4} $dig$ & $+$ & $Z_1$ & + &  \\ \hline
$Z_1$ & $zle$ & 0 & $zle'$ & \multirow{2}{*}{$\delta(zle,0) = (zle,0,left)$} \\ \cline{1-4} $zle$ & 0 & $Z_1$ & 0 &  \\ \hline 
$Z_1$ & $zle$ & 1 & $zle'$ & \multirow{2}{*}{$\delta(zle,1) = (zle,1,left)$} \\ \cline{1-4} $zle$ & 1 & $Z_1$ & 1 &  \\ \hline
$Z_1$ & $zle$ & $+$ & $zad'$ & \multirow{2}{*}{$\delta(zle,+) = (zad,+,left)$} \\ \cline{1-4} $zle$ & $+$ & $Z_1$ & + &  \\ \hline
$Z_1$ & $ole$ & 0 & $ole'$ & \multirow{2}{*}{$\delta(ole,0) = (ole,0,left)$} \\ \cline{1-4} $ole$ & 0 & $Z_1$ & 0 &  \\ \hline
$Z_1$ & $ole$ & 1 & $ole'$ & \multirow{2}{*}{$\delta(ole,1) = (ole,1,left)$} \\ \cline{1-4} $ole$ & 1 & $Z_1$ & 1 &  \\ \hline 
$Z_1$ & $ole$ & $+$ & $oad'$ & \multirow{2}{*}{$\delta(ole,+) = (oad,+,left)$} \\ \cline{1-4} $ole$ & $+$ & $Z_1$ & + &  \\ \hline
$Z_1$ & $zad$ & $y$ & $zad'$ & \multirow{2}{*}{$\delta(zad,y) = (zad,y,left)$} \\ \cline{1-4} $zad$ & $y$ & $Z_1$ & $y$ &  \\ \hline
$Z_1$ & $zad$ & $b$ & $zad'$ & \multirow{2}{*}{$\delta(zad,b) = (zad,b,left)$} \\ \cline{1-4} $zad$ & $b$ & $Z_1$ & $b$ &  \\ \hline
$Z_1$ & $zad$ & 0 & $y$ & \multirow{2}{*}{$\delta(zad,0) = (rig,y,right)$} \\ \cline{1-4} $zad$ & 0 & $Z_1$ & $rig$ &  \\ \hline
$Z_1$ & $zad$ & 1 & $b$ & \multirow{2}{*}{$\delta(zad,1) = (rig,b,right)$} \\ \cline{1-4} $zad$ & 1 & $Z_1$ & $rig$ &  \\ \hline
\end{tabular}
\end{table}

\begin{table}
\centering
\scriptsize
\begin{tabular}{|c|c|c|c|c|}
\multicolumn{3}{}{} & &  \\
\hline
$Z_1$ & $zad$ & = & $new'$ & \multirow{2}{*}{$\delta(zad,=) = (new,y,left)$} \\ \cline{1-4} $zad$ & = & $Z_1$ & $y$ &  \\ \hline
$Z_1$ & $oad$ & $y$ & $oad'$ & \multirow{2}{*}{$\delta(oad,y) = (oad,y,left)$} \\ \cline{1-4} $oad$ & $y$ & $Z_1$ & $y$ &  \\ \hline 
$Z_1$ & $oad$ & $b$ & $oad'$ & \multirow{2}{*}{$\delta(oad,b) = (oad,b,left)$} \\ \cline{1-4} $oad$ & $b$ & $Z_1$ & $b$ &  \\ \hline
$Z_1$ & $oad$ & 0 & $b$ & \multirow{2}{*}{$\delta(oad,0) = (rig,b,right)$} \\ \cline{1-4} $oad$ & 0 & $Z_1$ & $rig$ &  \\ \hline
$Z_1$ & $oad$ & 1 & $car'$ & \multirow{2}{*}{$\delta(oad,1) = (car,y,left)$} \\ \cline{1-4} $oad$ & 1 & $Z_1$ & $y$ &  \\ \hline
$Z_1$ & $oad$ & = & $new'$ & \multirow{2}{*}{$\delta(oad,=) = (new,b,left)$} \\ \cline{1-4} $oad$ & = & $Z_1$ & $b$ &  \\ \hline
$Z_1$ & $car$ & 0 & 1 & \multirow{2}{*}{$\delta(car,0) = (rig,1,right)$} \\ \cline{1-4} $car$ & 0 & $Z_1$ & $rig$ &  \\ \hline
$Z_1$ & $car$ & 1 & $car'$ & \multirow{2}{*}{$\delta(car,1) = (car,0,left)$} \\ \cline{1-4} $car$ & 1 & $Z_1$ & 0 &  \\ \hline
$Z_1$ & $car$ & = & $new'$ & \multirow{2}{*}{$\delta(car,=) = (new,1,left)$} \\ \cline{1-4} $car$ & = & $Z_1$ & 1 &  \\ \hline
$Z_1$ & $new$ & $B$ & = & \multirow{2}{*}{$\delta(new,B) = (rig,=,right)$} \\ \cline{1-4} $new$ & $B$ & $Z_1$ & $rig$ &  \\ \hline
$Z_1$ & $rig$ & 0 & 0 & \multirow{2}{*}{$\delta(rig,0) = (rig,0,right)$} \\ \cline{1-4} $rig$ & 0 & $Z_1$ & $rig$ &  \\ \hline
$Z_1$ & $rig$ & 1 & 1 & \multirow{2}{*}{$\delta(rig,1) = (rig,1,right)$} \\ \cline{1-4} $rig$ & 1 & $Z_1$ & $rig$ &  \\ \hline
$Z_1$ & $rig$ & $y$ & $y$ & \multirow{2}{*}{$\delta(rig,y) = (rig,y,right)$} \\ \cline{1-4} $rig$ & $y$ & $Z_1$ & $rig$ &  \\ \hline
$Z_1$ & $rig$ & $b$ & $b$ & \multirow{2}{*}{$\delta(rig,b) = (rig,b,right)$} \\ \cline{1-4} $rig$ & $b$ & $Z_1$ & $rig$ &  \\ \hline
$Z_1$ & $rig$ & + & + & \multirow{2}{*}{$\delta(rig,+) = (rig,+,right)$} \\ \cline{1-4} $rig$ & + & $Z_1$ & $rig$ &  \\ \hline
$Z_1$ & $rig$ & = & $dig'$ & \multirow{2}{*}{$\delta(rig,=) = (dig,B,left)$} \\ \cline{1-4} $rig$ & = & $Z_1$ & $B$ &  \\ \hline
$Z_1$ & $lef$ & 0 & $lef'$ & \multirow{2}{*}{$\delta(lef,0) = (lef,0,left)$} \\ \cline{1-4} $lef$ & 0 & $Z_1$ & 0 &  \\ \hline
$Z_1$ & $lef$ & 1 & $lef'$ & \multirow{2}{*}{$\delta(lef,1) = (lef,1,left)$} \\ \cline{1-4} $lef$ & 1 & $Z_1$ & 1 &  \\ \hline
$Z_1$ & $lef$ & $y$ & $lef'$ & \multirow{2}{*}{$\delta(lef,y) = (lef,y,left)$} \\ \cline{1-4} $lef$ & $y$ & $Z_1$ & $y$ &  \\ \hline
$Z_1$ & $lef$ & $b$ & $lef'$ & \multirow{2}{*}{$\delta(lef,b) = (lef,b,left)$} \\ \cline{1-4} $lef$ & $b$ & $Z_1$ & $b$ &  \\ \hline
$Z_1$ & $lef$ & = & $B$ & \multirow{2}{*}{$\delta(lef,=) = (fin,B,right)$} \\ \cline{1-4} $lef$ & = & $Z_1$ & $fin$ &  \\ \hline 
$Z_1$ & $fin$ & 0 & 0 & \multirow{2}{*}{$\delta(fin,0) = (fin,0,right)$} \\ \cline{1-4} $fin$ & 0 & $Z_1$ & $fin$ &  \\ \hline 
$Z_1$ & $fin$ & 1 & 1 & \multirow{2}{*}{$\delta(fin,1) = (fin,1,right)$} \\ \cline{1-4} $fin$ & 1 & $Z_1$ & $fin$ &  \\ \hline
$Z_1$ & $fin$ & $y$ & 0 & \multirow{2}{*}{$\delta(fin,y) = (fin,0,right)$} \\ \cline{1-4} $fin$ & $y$ & $Z_1$ & $fin$ &  \\ \hline
$Z_1$ & $fin$ & $b$ & 1 & \multirow{2}{*}{$\delta(fin,b) = (fin,1,right)$} \\ \cline{1-4} $fin$ & $b$ & $Z_1$ & $fin$ &  \\ \hline 
$Z_1$ & $fin$ & + & $B$ & \multirow{2}{*}{$\delta(fin,+) = (halt,B,right)$} \\ \cline{1-4} $fin$ & + & $Z_1$ & $halt$ &  \\ \hline
\end{tabular}
\end{table}

\begin{figure}
\centering
\includegraphics[width=1\textwidth]{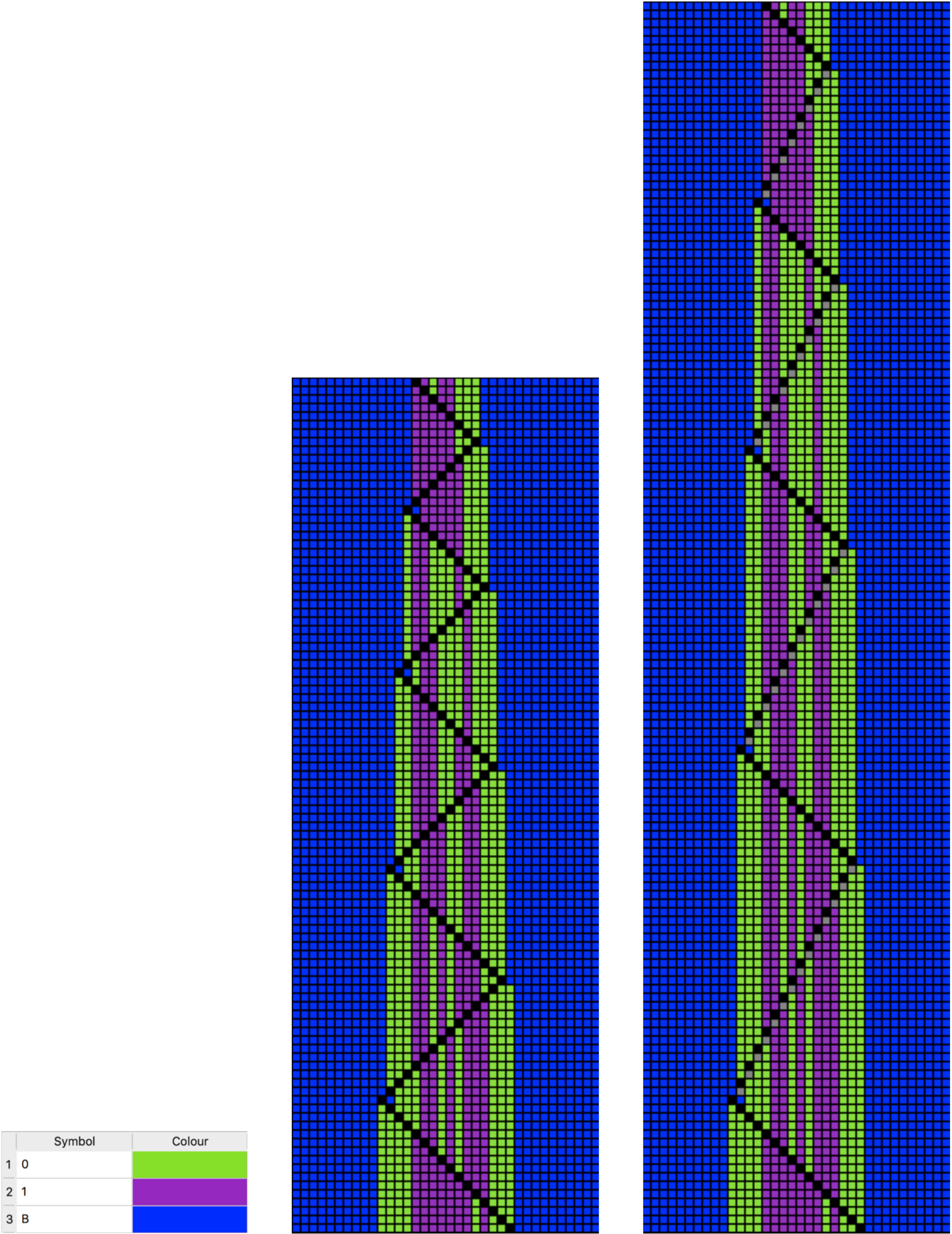}
\caption{Dynamical behaviour of the Turing machine $M$ and its equivalent cellular automaton $C_{R110}$ for the input string $101100$. (left) Representation of the symbols. (center) Turing machine dynamics. (right) Cellular automaton dynamics.}
\label{fig:exec05}
\end{figure}

The input string for both $M$ and $C_{BS}$ is $=00101+00101=$. The instantaneous description of $M$ when it stops is $01010B'halt'B$. The global configuration of $C_{BS}$ at the end of the simulation is $\dots B01010B'halt'B \dots$.

\subsection{Rule 110}

The second case of study is the famous elementary cellular automaton Rule 110, for details see \cite{kn:Mc99,kn:MMM06}.\footnote{Rule 110 repository. \url{http://uncomp.uwe.ac.uk/genaro/Rule110.html}} This rule is computationally universal simulating a cyclic tag system with a complicated system of gliders and collisions, for details see \cite{kn:Wolf02, kn:MMM11}.\footnote{Cyclic tag system working in Rule 110. \url{http://uncomp.uwe.ac.uk/genaro/rule110/ctsRule110.html}}

\begin{table}[th]
\centering
\caption{Turing machine transition table that simulate Rule 110.}
\label{cuad:func110}
\begin{tabular}{c|c|c|c}
state & 0 & 1 & $B$ \\
\hline
$S_{x0}$ & $S_{x0},0,R$ & $S_{01},1,R$ & $S_{B},0,L$ \\
$S_{01}$ & $S_{x0},1,R$ & $S_{11},1,R$ & \\
$S_{11}$ & $S_{x0},1,R$ & $S_{11},0,R$ & \\
$S_{B}$ & $S_{B},0,L$ & $S_{B},1,L$ & $S_{x0},0,R$ \\
\end{tabular}
\end{table}

Let the Turing machine be $M = (Q_M, \Sigma_M, \Gamma, \delta, q_0, B)$ capable of simulating the behaviour of the elementary cellular automaton Rule 110. This machine is one of several proposed with the same capacities described in Cook's paper \cite{kn:Cook04}. This way, $Q_M = \{S_{x0}, S_{01},$ $S_{11}, S_B\}$ is the set of states of the machine, $\Sigma_{M} = \{0,1\}$ is the set of input symbols, $\Gamma = \{0,1,B\}$ is the complete set of tape symbols, $\delta$ is the transition function shown in Table~\ref{cuad:func110}, $q_0 = a$ is the initial state of the machine and $B$ is the blank symbol.

\begin{table}[th]
\centering
\caption{Set of rules defined for $C_{R110}$ to simulate $M$.}
\label{cuad:specauto110}
\footnotesize
\begin{tabular}{|c|c|c|c|c|}
\hline
\multicolumn{3}{|c|}{$u(x)$} & $f(u(x))$ & equivalent $M$ transitions \\ \hline
        $Z_1$ & $Z_2$ & $S_{B}'$ & $S_{B}$ & - \\ \hline
        $Z_1$ & $S_{x0}$ & 0 & 0 & \multirow{2}{*}{$\delta(S_{x0},0) = (S_{x0},0,right)$} \\ \cline{1-4} $S_{x0}$ & 0 & $Z_1$ & $S_{x0}$ &  \\ \hline 
        $Z_1$ & $S_{x0}$ & 1 & 1 & \multirow{2}{*}{$\delta(S_{x0},1) = (S_{01},1,right)$} \\ \cline{1-4} $S_{x0}$ & 1 & $Z_1$ & $S_{01}$ &  \\ \hline 
        $Z_1$ & $S_{x0}$ & $B$ & $S_{B}'$ & \multirow{2}{*}{$\delta(S_{x0},B) = (S_{B},0,left)$} \\ \cline{1-4} $S_{x0}$ & $B$ & $Z_1$ & 0 &  \\ \hline
        $Z_1$ & $S_{01}$ & 0 & 1 & \multirow{2}{*}{$\delta(S_{01},0) = (S_{x0},1,right)$} \\ \cline{1-4} $S_{01}$ & 0 & $Z_1$ & $S_{x0}$ &  \\ \hline
        $Z_1$ & $S_{01}$ & 1 & 1 & \multirow{2}{*}{$\delta(S_{01},1) = (S_{11},1,right)$} \\ \cline{1-4} $S_{01}$ & 1 & $Z_1$ & $S_{11}$ &  \\ \hline 
        $Z_1$ & $S_{11}$ & 0 & 1 & \multirow{2}{*}{$\delta(S_{11},0) = (S_{x0},1,right)$} \\ \cline{1-4} $S_{11}$ & 0 & $Z_1$ & $S_{x0}$ &  \\ \hline 
        $Z_1$ & $S_{11}$ & 1 & 0 & \multirow{2}{*}{$\delta(S_{11},1) = (S_{11},0,right)$} \\ \cline{1-4} $S_{11}$ & 1 & $Z_1$ & $S_{11}$ &  \\ \hline 
        $Z_1$ & $S_{B}$ & 0 & $S_{B}'$ & \multirow{2}{*}{$\delta(S_{B},0) = (S_{B},0,left)$} \\ \cline{1-4} $S_{B}$ & 0 & $Z_1$ & 0 &  \\ \hline 
        $Z_1$ & $S_{B}$ & 1 & $S_{B}'$ & \multirow{2}{*}{$\delta(S_{B},1) = (S_{B},1,left)$} \\ \cline{1-4} $S_{B}$ & 1 & $Z_1$ & 1 &  \\ \hline 
        $Z_1$ & $S_{B}$ & $B$ & 0 & \multirow{2}{*}{$\delta(S_{B},B) = (S_{x0},0,right)$} \\ \cline{1-4} $S_{B}$ & $B$ & $Z_1$ & $S_{x0}$ &  \\ \hline	
\end{tabular}
\end{table}

Following the algorithm, we have the whole set of rules defined for the cellular automaton $C_{R110}$ to simulate $M$ in Table~\ref{cuad:func110}.

The input string for both $M$ and $C_{R110}$ machines is $1011000$. The instantaneous description of $M$ when it stops is $00001111$ $1110100S_{x0}B$ (Fig.~\ref{fig:exec05} center). The global configuration of $C_{R110}$ at the end of the simulation is $\dots B00001111111010$ $0S_{x0}B \dots$ (Fig.~\ref{fig:exec05} right).

We will note that in this system is not relevant a halt condition because the goal is yield the belong string. In a cellular space for Rule 110 starting with the string $1011$, it yields the string $111111101$ five generations later.

\subsection{Reversible Turing machine}

\begin{figure}
\centering
\includegraphics[width=0.36\textwidth]{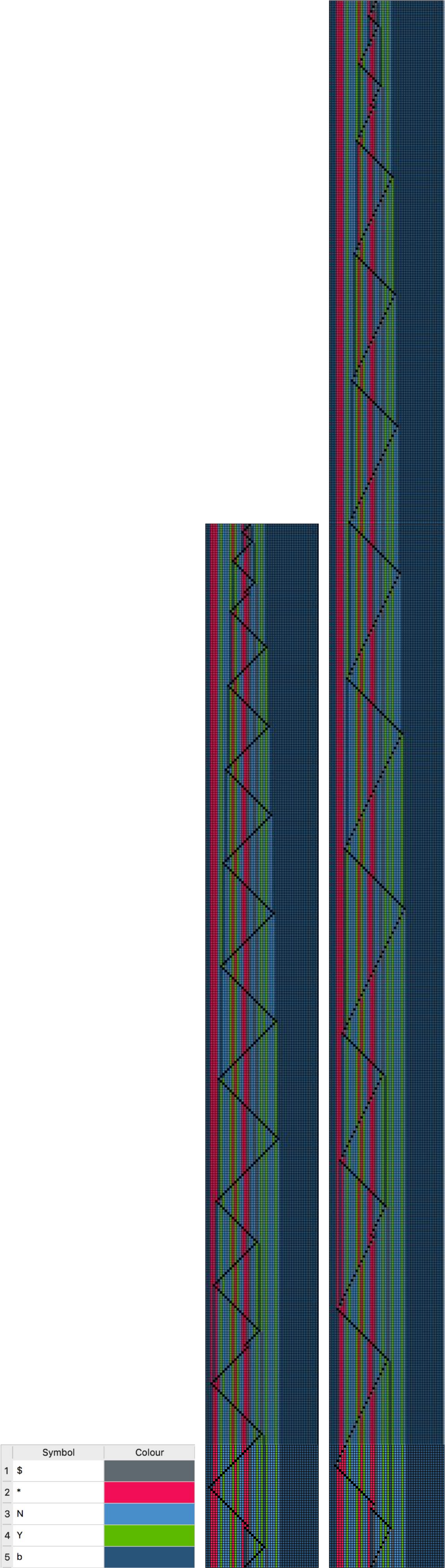}
\caption{Dynamical behaviour of the Turing machine $M$ and its equivalent cellular automaton $C_{UR}$ for the input string $\ast$$\ast$$\ast$$NYNNNY*YNN*b\$NNYNYN$. (left) Representation of the symbols. (center) Turing machine dynamics. (right) Cellular automaton dynamics.}
\label{fig:exec06}
\end{figure}

In our last case, we consider a special kind of Turing machines that are reversible in its table transitions. It is an universal reversible Turing machine able to simulate any cyclic tag system proposed by Morita and Yamaguchi with 17 states and five symbols. For details see \cite{kn:MY07}.

Let the Turing machine be $M = (Q_M, \Sigma_M, \Gamma, \delta, q_0, B)$ which is an universal reversible Turing machine. $Q_M = \{q_0, q_1, q_2, \dots, q_{16}, halt\}$ is the set of states of the machine, $\Sigma_M = \{Y, N, *, \$\}$ is the set of input symbols, $\Gamma = \{Y, N, *, \$, b\}$ is the complete set of tape symbols, $\delta$ is the transition function shown in the Table \ref{cuad:urtm}, $q_0$ is the initial state of the machine and $b$ is the blank symbol.

\begin{table}[th]
\centering
\caption{Transition table for a universal reversible Turing machine $(17,5)$.}
\label{cuad:urtm}
\footnotesize
\begin{tabular}{c|c|c|c|c|c}
        State & $b$ & $Y$ & $N$ & * & \$ \\ \hline
        $q_{0}$ & $q_2,\$,L$ & $q_1,\$,L$ & $q_{13},b,L$ & & \\
        $q_{1}$ &  & $q_1,Y,L$ & $q_1,N,L$ & $q_0,*,R$ & $q_1,b,L$ \\
        $q_{2}$ & $q_3,*,L$ & $q_2,Y,L$ & $q_2,N,L$ & $q_2,*,L$ &  \\
        $q_{3}$ & $q_{12},b,R$ & $q_4,b,R$ & $q_7,b,R$ & $q_{10},b,R$ &  \\
        $q_{4}$ & $q_5,Y,R$ & $q_4,Y,R$ & $q_4,N,R$ & $q_4,*,R$ & $q_4,\$,R$ \\
        $q_{5}$ & $q_6,b,L$ &  &  &  &  \\
        $q_{6}$ & $q_3,Y,L$ & $q_6,Y,L$ & $q_6,N,L$ & $q_6,*,L$ & $q_6,\$,L$ \\
        $q_{7}$ & $q_8,N,R$ & $q_7,Y,R$ & $q_7,N,R$ & $q_7,*,R$ & $q_7,\$,R$ \\
        $q_{8}$ & $q_9,b,L$ &  &  &  &  \\
        $q_{9}$ & $q_3,N,L$ & $q_9,Y,L$ & $q_9,N,L$ & $q_9,*,L$ & $q_9,\$,L$ \\
        $q_{10}$ &  & $q_{10},Y,R$ & $q_{10},N,R$ & $q_{10},*,R$ & $q_{11},\$,R$ \\
        $q_{11}$ &  & $q_{11},Y,R$ & $q_{11},N,R$ & $q_{11},*,R$ & $q_0,Y,R$ \\
        $q_{12}$ &  & $q_{12},Y,R$ & $q_{12},N,R$ & $q_{12},*,R$ & $q_3,\$,L$ \\
        $q_{13}$ & $q_{14},*,L$ & $q_{13},Y,L$ & $q_{13},N,L$ & $q_{13},*,L$ & $q_{13},\$,L$ \\
        $q_{14}$ & $q_{16},b,R$ & $q_{14},Y,L$ & $q_{14},N,L$ & $q_{15},b,R$ &  \\
        $q_{15}$ & $q_0,N,R$ & $q_{15},Y,R$ & $q_{15},N,R$ & $q_{15},*,R$ & $q_{15},\$,R$ \\
        $q_{16}$ &  & $q_{16},Y,R$ & $q_{16},N,R$ & $q_{16},*,R$ & $q_{14},\$,L$ \\
\end{tabular}
\end{table}

As we did in the last two examples, we can observe that both, the behaviour of the Turing machine $M$ (Fig.~\ref{fig:exec06} center) and the behaviour of its equivalent cellular automaton $C_{UR}$ (Fig.~\ref{fig:exec06} right) differ only in the amount of time required to process the movements of the head to the left. The set of rules defined for $C_{UR}$ is shown in Table~\ref{cuad:specautoURTM}.

\begin{table}
\centering
\caption{Set of rules defined for $C_{UR}$ to simulate $M$.}
\label{cuad:specautoURTM}
\scriptsize
\begin{tabular}{|c|c|c|c|c|}
    	\hline
    	\multicolumn{3}{|c|}{$u(x)$} & $f(u(x))$ & equivalent $M$ transitions \\ \hline
        $Z_1$ & $Z_2$ & $q_{1}'$ & $q_{1}$ & - \\ \hline
        $Z_1$ & $Z_2$ & $q_{2}'$ & $q_{2}$ & - \\ \hline
        $Z_1$ & $Z_2$ & $q_{3}'$ & $q_{3}$ & - \\ \hline
        $Z_1$ & $Z_2$ & $q_{6}'$ & $q_{6}$ & - \\ \hline
        $Z_1$ & $Z_2$ & $q_{9}'$ & $q_{9}$ & - \\ \hline
        $Z_1$ & $Z_2$ & $q_{13}'$ & $q_{13}$ & - \\ \hline
        $Z_1$ & $Z_2$ & $q_{14}'$ & $q_{14}$ & - \\ \hline
        $Z_1$ & $q_0$ & $b$ & $q_2'$ & \multirow{2}{*}{$\delta(q_0,b) = (q_2,\$,Left)$} \\ \cline{1-4} $q_0$ & $b$ & $Z_1$ & \$ &  \\ \hline 
        $Z_1$ & $q_0$ & $Y$ & $q_1'$ & \multirow{2}{*}{$\delta(q_0,Y) = (q_1,\$,Left)$} \\ \cline{1-4} $q_0$ & $Y$ & $Z_1$ & \$ &  \\ \hline 
        $Z_1$ & $q_0$ & $N$ & $q_{13}'$ & \multirow{2}{*}{$\delta(q_0,N) = (q_{13},b,Left)$} \\ \cline{1-4} $q_0$ & $N$ & $Z_1$ & $b$ &  \\ \hline 
        $Z_1$ & $q_1$ & $Y$ & $q_1'$ & \multirow{2}{*}{$\delta(q_1,Y) = (q_1,Y,Left)$} \\ \cline{1-4} $q_1$ & $Y$ & $Z_1$ & $Y$ &  \\ \hline 
        $Z_1$ & $q_1$ & $N$ & $q_1'$ & \multirow{2}{*}{$\delta(q_1,N) = (q_1,N,Left)$} \\ \cline{1-4} $q_1$ & $N$ & $Z_1$ & $N$ &  \\ \hline 
        $Z_1$ & $q_1$ & * & * & \multirow{2}{*}{$\delta(q_1,*) = (q_0,*,Right)$} \\ \cline{1-4} $q_1$ & * & $Z_1$ & $q_0$ &  \\ \hline 
        $Z_1$ & $q_1$ & \$ & $q_1'$ & \multirow{2}{*}{$\delta(q_1,\$) = (q_1,b,Left)$} \\ \cline{1-4} $q_1$ & \$ & $Z_1$ & $b$ &  \\ \hline 
        $Z_1$ & $q_2$ & $b$ & $q_3'$ & \multirow{2}{*}{$\delta(q_2,b) = (q_3,*,Left)$} \\ \cline{1-4} $q_2$ & $b$ & $Z_1$ & * &  \\ \hline 
        $Z_1$ & $q_2$ & $Y$ & $q_2'$ & \multirow{2}{*}{$\delta(q_2,Y) = (q_2,Y,Left)$} \\ \cline{1-4} $q_2$ & $Y$ & $Z_1$ & $Y$ &  \\ \hline 
        $Z_1$ & $q_2$ & $N$ & $q_2'$ & \multirow{2}{*}{$\delta(q_2,N) = (q_2,N,Left)$} \\ \cline{1-4} $q_2$ & $N$ & $Z_1$ & $N$ &  \\ \hline 
        $Z_1$ & $q_2$ & * & $q_2'$ & \multirow{2}{*}{$\delta(q_2,*) = (q_2,*,Left)$} \\ \cline{1-4} $q_2$ & * & $Z_1$ & * &  \\ \hline 
        $Z_1$ & $q_3$ & $b$ & $b$ & \multirow{2}{*}{$\delta(q_3,b) = (q_{12},b,Right)$} \\ \cline{1-4} $q_3$ & $b$ & $Z_1$ & $q_{12}$ &  \\ \hline 
        $Z_1$ & $q_3$ & $Y$ & $b$ & \multirow{2}{*}{$\delta(q_3,Y) = (q_4,b,Right)$} \\ \cline{1-4} $q_3$ & $Y$ & $Z_1$ & $q_4$ &  \\ \hline 
        $Z_1$ & $q_3$ & $N$ & $b$ & \multirow{2}{*}{$\delta(q_3,N) = (q_7,b,Right)$} \\ \cline{1-4} $q_3$ & $N$ & $Z_1$ & $q_7$ &  \\ \hline 
        $Z_1$ & $q_3$ & * & $b$ & \multirow{2}{*}{$\delta(q_3,*) = (q_{10},b,Right)$} \\ \cline{1-4} $q_3$ & * & $Z_1$ & $q_{10}$ &  \\ \hline 
        $Z_1$ & $q_4$ & $b$ & $Y$ & \multirow{2}{*}{$\delta(q_4,b) = (q_5,Y,Right)$} \\ \cline{1-4} $q_4$ & $b$ & $Z_1$ & $q_5$ &  \\ \hline 
        $Z_1$ & $q_4$ & $Y$ & $Y$ & \multirow{2}{*}{$\delta(q_4,Y) = (q_4,Y,Right)$} \\ \cline{1-4} $q_4$ & $Y$ & $Z_1$ & $q_4$ &  \\ \hline 
        $Z_1$ & $q_4$ & $N$ & $N$ & \multirow{2}{*}{$\delta(q_4,N) = (q_4,N,Right)$} \\ \cline{1-4} $q_4$ & $N$ & $Z_1$ & $q_4$ &  \\ \hline 
        $Z_1$ & $q_4$ & * & * & \multirow{2}{*}{$\delta(q_4,*) = (q_4,*,Right)$} \\ \cline{1-4} $q_4$ & * & $Z_1$ & $q_4$ &  \\ \hline 
        $Z_1$ & $q_4$ & \$ & \$ & \multirow{2}{*}{$\delta(q_4,\$) = (q_4,\$,Right)$} \\ \cline{1-4} $q_4$ & \$ & $Z_1$ & $q_4$ &  \\ \hline 
        $Z_1$ & $q_5$ & $b$ & $q_6'$ & \multirow{2}{*}{$\delta(q_5,b) = (q_6,b,Left)$} \\ \cline{1-4} $q_5$ & $b$ & $Z_1$ & $b$ &  \\ \hline 
\end{tabular}
\end{table}
        
\begin{table}
\centering
\scriptsize
\begin{tabular}{|c|c|c|c|c|}
\multicolumn{3}{}{} & &  \\
\hline
        $Z_1$ & $q_6$ & $b$ & $q_3'$ & \multirow{2}{*}{$\delta(q_6,b) = (q_3,Y,Left)$} \\ \cline{1-4} $q_6$ & $b$ & $Z_1$ & $Y$ &  \\ \hline 
        $Z_1$ & $q_6$ & $Y$ & $q_6'$ & \multirow{2}{*}{$\delta(q_6,Y) = (q_6,Y,Left)$} \\ \cline{1-4} $q_6$ & $Y$ & $Z_1$ & $Y$ &  \\ \hline 
        $Z_1$ & $q_6$ & $N$ & $q_6'$ & \multirow{2}{*}{$\delta(q_6,N) = (q_6,N,Left)$} \\ \cline{1-4} $q_6$ & $N$ & $Z_1$ & $N$ &  \\ \hline 
        $Z_1$ & $q_6$ & * & $q_6'$ & \multirow{2}{*}{$\delta(q_6,*) = (q_6,*,Left)$} \\ \cline{1-4} $q_6$ & * & $Z_1$ & * &  \\ \hline 
        $Z_1$ & $q_6$ & \$ & $q_6'$ & \multirow{2}{*}{$\delta(q_6,\$) = (q_6,\$,Left)$} \\ \cline{1-4} $q_6$ & \$ & $Z_1$ & \$ &  \\ \hline 
        $Z_1$ & $q_7$ & $b$ & $N$ & \multirow{2}{*}{$\delta(q_7,b) = (q_8,N,Right)$} \\ \cline{1-4} $q_7$ & $b$ & $Z_1$ & $q_8$ &  \\ \hline 
        $Z_1$ & $q_7$ & $Y$ & $Y$ & \multirow{2}{*}{$\delta(q_7,Y) = (q_7,Y,Right)$} \\ \cline{1-4} $q_7$ & $Y$ & $Z_1$ & $q_7$ &  \\ \hline 
        $Z_1$ & $q_7$ & $N$ & $N$ & \multirow{2}{*}{$\delta(q_7,N) = (q_7,N,Right)$} \\ \cline{1-4} $q_7$ & $N$ & $Z_1$ & $q_7$ &  \\ \hline 
        $Z_1$ & $q_7$ & * & * & \multirow{2}{*}{$\delta(q_7,*) = (q_7,*,Right)$} \\ \cline{1-4} $q_7$ & * & $Z_1$ & $q_7$ &  \\ \hline 
        $Z_1$ & $q_7$ & \$ & \$ & \multirow{2}{*}{$\delta(q_7,\$) = (q_7,\$,Right)$} \\ \cline{1-4} $q_7$ & \$ & $Z_1$ & $q_7$ &  \\ \hline 
        $Z_1$ & $q_8$ & $b$ & $q_9'$ & \multirow{2}{*}{$\delta(q_8,b) = (q_9,b,Left)$} \\ \cline{1-4} $q_8$ & $b$ & $Z_1$ & $b$ &  \\ \hline 
        $Z_1$ & $q_9$ & $b$ & $q_3'$ & \multirow{2}{*}{$\delta(q_9,b) = (q_3,N,Left)$} \\ \cline{1-4} $q_9$ & $b$ & $Z_1$ & $N$ &  \\ \hline 
        $Z_1$ & $q_9$ & $Y$ & $q_9'$ & \multirow{2}{*}{$\delta(q_9,Y) = (q_9,Y,Left)$} \\ \cline{1-4} $q_9$ & $Y$ & $Z_1$ & $Y$ &  \\ \hline 
        $Z_1$ & $q_9$ & $N$ & $q_9'$ & \multirow{2}{*}{$\delta(q_9,N) = (q_9,N,Left)$} \\ \cline{1-4} $q_9$ & $N$ & $Z_1$ & $N$ &  \\ \hline 
        $Z_1$ & $q_9$ & * & $q_9'$ & \multirow{2}{*}{$\delta(q_9,*) = (q_9,*,Left)$} \\ \cline{1-4} $q_9$ & * & $Z_1$ & * &  \\ \hline 
        $Z_1$ & $q_9$ & \$ & $q_9'$ & \multirow{2}{*}{$\delta(q_9,\$) = (q_9,\$,Left)$} \\ \cline{1-4} $q_9$ & \$ & $Z_1$ & \$ &  \\ \hline
        $Z_1$ & $q_{10}$ & $Y$ & $Y$ & \multirow{2}{*}{$\delta(q_{10},Y) = (q_{10},Y,Right)$} \\ \cline{1-4} $q_{10}$ & $Y$ & $Z_1$ & $q_{10}$ &  \\ \hline 
        $Z_1$ & $q_{10}$ & $N$ & $N$ & \multirow{2}{*}{$\delta(q_{10},N) = (q_{10},N,Right)$} \\ \cline{1-4} $q_{10}$ & $N$ & $Z_1$ & $q_{10}$ &  \\ \hline 
        $Z_1$ & $q_{10}$ & * & * & \multirow{2}{*}{$\delta(q_{10},*) = (q_{10},*,Right)$} \\ \cline{1-4} $q_{10}$ & * & $Z_1$ & $q_{10}$ &  \\ \hline 
        $Z_1$ & $q_{10}$ & \$ & \$ & \multirow{2}{*}{$\delta(q_{10},\$) = (q_{11},\$,Right)$} \\ \cline{1-4} $q_{10}$ & \$ & $Z_1$ & $q_{11}$ &  \\ \hline 
        $Z_1$ & $q_{11}$ & $Y$ & $Y$ & \multirow{2}{*}{$\delta(q_{11},Y) = (q_{11},Y,Right)$} \\ \cline{1-4} $q_{11}$ & $Y$ & $Z_1$ & $q_{11}$ &  \\ \hline 
        $Z_1$ & $q_{11}$ & $N$ & $N$ & \multirow{2}{*}{$\delta(q_{11},N) = (q_{11},N,Right)$} \\ \cline{1-4} $q_{11}$ & $N$ & $Z_1$ & $q_{11}$ &  \\ \hline 
        $Z_1$ & $q_{11}$ & * & * & \multirow{2}{*}{$\delta(q_{11},*) = (q_{11},*,Right)$} \\ \cline{1-4} $q_{11}$ & * & $Z_1$ & $q_{11}$ &  \\ \hline 
        $Z_1$ & $q_{11}$ & \$ & $Y$ & \multirow{2}{*}{$\delta(q_{11},\$) = (q_0,Y,Right)$} \\ \cline{1-4} $q_{11}$ & \$ & $Z_1$ & $q_0$ &  \\ \hline 
        $Z_1$ & $q_{12}$ & $Y$ & $Y$ & \multirow{2}{*}{$\delta(q_{12},Y) = (q_{12},Y,Right)$} \\ \cline{1-4} $q_{12}$ & $Y$ & $Z_1$ & $q_{12}$ &  \\ \hline 
        $Z_1$ & $q_{12}$ & $N$ & $N$ & \multirow{2}{*}{$\delta(q_{12},N) = (q_{12},N,Right)$} \\ \cline{1-4} $q_{12}$ & $N$ & $Z_1$ & $q_{12}$ &  \\ \hline 
        $Z_1$ & $q_{12}$ & * & * & \multirow{2}{*}{$\delta(q_{12},*) = (q_{12},*,Right)$} \\ \cline{1-4} $q_{12}$ & * & $Z_1$ & $q_{12}$ &  \\ \hline 
        $Z_1$ & $q_{12}$ & \$ & $q_3'$ & \multirow{2}{*}{$\delta(q_{12},\$) = (q_3,\$,Left)$} \\ \cline{1-4} $q_{12}$ & \$ & $Z_1$ & \$ &  \\ \hline 
\end{tabular}
\end{table}
        
\begin{table}
\centering
\scriptsize
\begin{tabular}{|c|c|c|c|c|}
\multicolumn{3}{}{} & &  \\
\hline
        $Z_1$ & $q_{13}$ & $b$ & $q_{14}'$ & \multirow{2}{*}{$\delta(q_{13},b) = (q_{14},*,Left)$} \\ \cline{1-4} $q_{13}$ & $b$ & $Z_1$ & * &  \\ \hline 
        $Z_1$ & $q_{13}$ & $Y$ & $q_{13}'$ & \multirow{2}{*}{$\delta(q_{13},Y) = (q_{13},Y,Left)$} \\ \cline{1-4} $q_{13}$ & $Y$ & $Z_1$ & $Y$ &  \\ \hline 
        $Z_1$ & $q_{13}$ & $N$ & $q_{13}'$ & \multirow{2}{*}{$\delta(q_{13},N) = (q_{13},N,Left)$} \\ \cline{1-4} $q_{13}$ & $N$ & $Z_1$ & $N$ &  \\ \hline 
        $Z_1$ & $q_{13}$ & * & $q_{13}'$ & \multirow{2}{*}{$\delta(q_{13},*) = (q_{13},*,Left)$} \\ \cline{1-4} $q_{13}$ & * & $Z_1$ & * &  \\ \hline 
        $Z_1$ & $q_{13}$ & \$ & $q_{13}'$ & \multirow{2}{*}{$\delta(q_{13},\$) = (q_{13},\$,Left)$} \\ \cline{1-4} $q_{13}$ & \$ & $Z_1$ & \$ &  \\ \hline
        $Z_1$ & $q_{14}$ & $b$ & $b$ & \multirow{2}{*}{$\delta(q_{14},b) = (q_{15},b,Right)$} \\ \cline{1-4} $q_{14}$ & $b$ & $Z_1$ & $q_{15}$ &  \\ \hline
        $Z_1$ & $q_{14}$ & $Y$ & $q_{14}'$ & \multirow{2}{*}{$\delta(q_{14},Y) = (q_{14},Y,Left)$} \\ \cline{1-4} $q_{14}$ & $Y$ & $Z_1$ & $Y$ &  \\ \hline        
        $Z_1$ & $q_{14}$ & $N$ & $q_{14}'$ & \multirow{2}{*}{$\delta(q_{14},N) = (q_{14},N,Left)$} \\ \cline{1-4} $q_{14}$ & $N$ & $Z_1$ & $N$ &  \\ \hline
        $Z_1$ & $q_{14}$ & * & $b$ & \multirow{2}{*}{$\delta(q_{14},*) = (q_{14},b,Right)$} \\ \cline{1-4} $q_{14}$ & * & $Z_1$ & $q_{14}$ &  \\ \hline
        $Z_1$ & $q_{15}$ & $b$ & $N$ & \multirow{2}{*}{$\delta(q_{15},b) = (q_0,N,Right)$} \\ \cline{1-4} $q_{15}$ & $b$ & $Z_1$ & $q_0$ &  \\ \hline
        $Z_1$ & $q_{15}$ & $Y$ & $Y$ & \multirow{2}{*}{$\delta(q_{15},Y) = (q_{15},Y,Right)$} \\ \cline{1-4} $q_{15}$ & $Y$ & $Z_1$ & $q_{15}$ &  \\ \hline
        $Z_1$ & $q_{15}$ & $N$ & $N$ & \multirow{2}{*}{$\delta(q_{15},N) = (q_{15},N,Right)$} \\ \cline{1-4} $q_{15}$ & $N$ & $Z_1$ & $q_{15}$ &  \\ \hline
        $Z_1$ & $q_{15}$ & * & * & \multirow{2}{*}{$\delta(q_{15},*) = (q_{15},*,Right)$} \\ \cline{1-4} $q_{15}$ & * & $Z_1$ & $q_{15}$ &  \\ \hline
        $Z_1$ & $q_{15}$ & \$ & \$ & \multirow{2}{*}{$\delta(q_{15},\$) = (q_{15},\$,Right)$} \\ \cline{1-4} $q_{15}$ & \$ & $Z_1$ & $q_{15}$ &  \\ \hline
        $Z_1$ & $q_{16}$ & $Y$ & $Y$ & \multirow{2}{*}{$\delta(q_{16},Y) = (q_{16},Y,Right)$} \\ \cline{1-4} $q_{16}$ & $Y$ & $Z_1$ & $q_{16}$ &  \\ \hline
        $Z_1$ & $q_{16}$ & $N$ & $N$ & \multirow{2}{*}{$\delta(q_{16},N) = (q_{16},N,Right)$} \\ \cline{1-4} $q_{16}$ & $N$ & $Z_1$ & $q_{16}$ &  \\ \hline
        $Z_1$ & $q_{16}$ & * & * & \multirow{2}{*}{$\delta(q_{16},*) = (q_{16},*,Right)$} \\ \cline{1-4} $q_{16}$ & * & $Z_1$ & $q_{16}$ &  \\ \hline
        $Z_1$ & $q_{16}$ & \$ & $q_{14}'$ & \multirow{2}{*}{$\delta(q_{16},\$) = (q_{14},\$,Left)$} \\ \cline{1-4} $q_{16}$ & \$ & $Z_1$ & \$ &  \\ \hline 
\end{tabular}
\end{table}

The input string for both $M$ and $C_{UR}$ is $***NYNNNY*YNN*b\$N$ $NYNYN$. The instantaneous description of $M$ when it stops is $***NYNN$ $NY*YNN*q_{1}bbNNYNYN\$NNNYN$. The global configuration of $C_{UR}$ at the end of the simulation is $\dots b***NYNNNY*YNN*q_{1}bbNNY$ $NYN\$NNNYNb\dots$.

\section{Conclusions}

We proved that  one-dimensional cellular automata $C_{BS}$, $C_{R110}$, and $C_{UR}$ are universal, simulating dynamics of Turing machines and exploring the spacial dynamics of conventional Turing machines in two dimensions such as cellular automata evolutions. This way, the algorithm proposed in this paper may convert any Turing machine to an equivalent one-dimensional cellular automaton.

The software developed (in Java) and used to calculate all simulations in the paper is free and available from the next link: \url{http://uncomp.uwe.ac.uk/genaro/Papers/Thesis_files/maquinaTuring.java.tar.gz}.


\end{document}